# Green luminescence and calculated optical properties of Cu ions in ZnO

O. Volnianska and P. Bogusławski[*]

*Institute of Physics PAS, 02-668 Warsaw, Poland*

There are two characteristic optical transitions associated with Cu ions in ZnO, the 0.72 eV infrared and the 2.86 eV structured green luminescence (SGL). While the former is unambiguously related with the $d(Cu^{2+})$ intra-shell transition, there is no generally accepted mechanism of the SGL. Besides the original model of Dingle [Phys. Rev. Lett. 23, 579 (1969)], two other mechanisms were recently proposed. We report an analysis of the optical properties of Cu in ZnO by ab initio calculations. The GGA+$U$ approach is used, with the +$U$ corrections applied to $d(Zn)$, $p(O)$ and $d(Cu)$ orbitals. The results, compared with the available experimental data, support the Dingle's model, in which the SGL originates in the ($Cu^{1+}$, hole) $\rightarrow Cu^{2+}$ transition. A good agreement with experiment is obtained also for the internal transition at 0.72 eV. The absence of an expected radiative transition at about 2 eV is explained by its quasi-forbidden character.

## 1. INTRODUCTION

Thanks to its wide band gap and large exciton binding energy, zinc oxide attracts a considerable attention as optoelectronic device material. Applications of ZnO include the source of blue or UV light, solar cells, or phosphorescent thin films [1-4]. In spite of a long-standing effort, $p$-doping of ZnO still awaits for a fully satisfactory solution. Among various dopants, group IB ions, and in particular Cu, was examined. The acceptor level of copper is situated about 3 eV above the valence band maximum (VBM), which makes Cu non-practical for achieving $p$-conductivity. Experimental studies of Cu ions in ZnO were extended recently to ZnO quantum dots [5-10]. The interest in Cu:ZnO increased after the discovery of the Cu-related room temperature ferromagnetism [11-16]. This topic is out of scope of our paper, but the correct description of the electronic structure of Cu ions is prerequisite for the correct explanation of the Cu-Cu magnetic coupling.

Regarding the optical properties, two types of green luminescence of Cu-doped ZnO are observed. The first one, often activated by annealing, is characterized by sharp series of phonon replicas of the 2.86 eV zero-phonon line (ZPL). It is called the structured green luminescence, SGL [17-33]. The second type consists in a broad line centred at about 2.4 eV, and it is not analysed here. There is little doubt that in all experiments reporting the SGL, the 2.86 eV line must be due to the same transition involving the same center, Cu, in the same charge state [34]. In spite of the considerable effort lasting since five decades, there is no generally accepted identification of the origin of the SGL. The commonly evoked mechanism was proposed by Dingle [18], and consists in the capture of a photoelectron by $Cu^{2+}$, followed by the recombination $Cu^{1+} \rightarrow Cu^{2+}$ involving a free hole.

In addition to the SGL, the photoluminescence (PL) line at 0.72 eV was observed. In this case, there is a consensus regarding its origin, and the luminescence is ascribed to an internal $Cu^{2+}$ transition [17]. One can observe that the occurrence of this transition implies that there are gap two levels of $Cu^{2+}$. Indeed, previous ab initio calculations, albeit focused on the SGL transition [33, 35-38], have demonstrated that there are two levels of $Cu^{2+}$ in the band gap [36, 37]. Thus, the internal transition between these levels can take place, in agreement with experiment, but the process was not considered in the previous theoretical works. Moreover, the presence of two Cu levels in the band gap suggests that more than one charge transfer transition can be expected, but only one such transition, the SGL, was observed, a fact which also requires an explanation.

Here, we aim at explaining the observed luminescence properties of Cu:ZnO. To some extent, the confusion regarding the mechanism of the SGL is due to the fact that the available rich experimental results are not always considered exhaustively, and some proposed models are in conflict with the established data. In this situation we begin by a summary of the experimental findings.

(i) As it was revealed by parallel optical and electron paramagnetic resonance (EPR) measurements [18], the ground state of SGL is a copper ion in the $Cu^{2+}$ charge state. The zero-phonon line (ZPL) energy is 2.86 eV. Accordingly, the PL excitation spectrum shows an offset at 2.9 eV [29]. The ZPL is split by about 0.1 meV by the isotope effect [18, 28, 31]. The phonon replicas match very well the ZnO LO phonon energy of 72 meV.

(ii) Angular dependence of the EPR line reveals that both the ground and the excited state of Cu have the $C_{3v}$ point symmetry [17, 18]. This finding sets limits on the possible symmetries of the Jahn-Teller distortions. Indeed, $C_{3v}$ implies that the symmetry of the atomic configuration of Cu is the same as that of a host atom, with three equivalent in-plane (or basal) bonds that can differ from the fourth one oriented along the $c$-axis. Thus, the $C_{3v}$ symmetry characterizes both $d^9$ and $d^{10}$ electronic configuration assumed by $Cu^{2+}$ and $Cu^+$, respectively, even including the Jahn-Teller distortion of $Cu^{2+}$. Importantly, this result demonstrates that the local symmetry of the $Cu^{2+}$ ions corresponds to isolated centers, *i.e.*, there are no defects and/or dopants in the nearest Cu surrounding.

(iii) In some papers, three closely spaced ZPLs were observed [19, 20]. They are separated by about 10 meV, and give rise to three series of LO phonon replicas. The interpretation of this effect relies on the fact that the VBM in ZnO is split into three levels, $\Gamma_7$, $\Gamma_9$, and $\Gamma_7$, by the combined effects of the wurtzite crystal field and the spin-orbit coupling. In more recent works [24-26, 30, 33], two rather than three ZPL lines were observed. They are separated by about 20-30 meV. It is tempting to assume that both the two-fold and the three-fold splitting are due to the same physical factor, *i.e.*, the splitting of the VBM. This can differ from sample to sample because of the actual strain conditions in ZnO layers. Indeed, the VBM splitting in ZnO is small, and thus the exact splitting energies of the VBM are sensitive to possible strains present in ZnO layers.

An alternative interpretation of the ZPL splitting was proposed in Ref. [22], according to which the SGL is a donor-acceptor pair (DAP) recombination, and that the two observed sets of lines originate in the presence of two shallow donors, at 30 and 60 meV. However, there are two problems regarding this interpretation. The first one is the sharpness of the 2.86 eV line: typically, the DAP lines are broadened by the fact that the donor-acceptor distance changes from pair to pair. Second, the results quoted in the point (ii) demonstrated that the SGL is due to an isolated center, and not to a defect pair.

(iv) Time resolved experiments provided an important insight into the mechanism of the SGL. In particular, Xing *et al.* [27] used the above band gap excitation and measured time decays of both band edge PL and that of the SGL. The results demonstrated that the SGL is preceded by the charge transfer process of photoelectrons from the conduction band minimum (CBM) to $Cu^{2+}$, leading to formation of $Cu^+$ state. The electron capture by $Cu^{2+}$ severely limits the band gap emission of Cu:ZnO, because the capture time is shorter than the decay time of radiative recombination in pure ZnO. In fact, the decay time of the band gap emission in Cu:ZnO is fully correlated with the rise time of the SGL. Also, $Cu^{2+}$ ions capture electrons rather than holes [27]. The temporal dependence of the SGL is qualitatively different when above band gap and below band gap excitations are used [33]. In the latter case, besides the fast SGL there is a second SGL channel with decay times of the order of minutes [33].

(v) The second characteristic emission related with Cu occurs at about 0.72 eV [17, 20, 23, 28], with the line split by isotope effect [23, 28]. According to the commonly accepted interpretation, the 0.72 eV line originates in the internal transition of $Cu^{2+}(d^9)$, between $t_{2\downarrow}$ and $e_\downarrow$ levels, *i.e.*, between $^2T_2$ and $^2E$ configurations. It was observed also in absorption [23, 28]. The 0.72 eV luminescence was extensively analysed in Ref. [17]. This transition is strong in ZnS, but not in ZnO, for reasons explained in Ref. [20]. We stress that while only a few experimental reports on this line are available, from the point of view of the electronic structure theory a correct description of this line is as important as that of the SGL.

Summarizing, the SGL typically occurs after the above band gap photoexcitation. Photoelectrons are captured by $Cu^{2+}$ ions, and then recombination with free holes takes place. The 2.86 eV ZPL line is fine split into two or three lines distant by 10-20 meV, which most probably reflects the splitting of the VBM. $Cu^{2+}$ centers participating in SGL are isolated, and their local symmetry is $C_{3v}$. Besides SGL, there is an intracenter transition at 0.72 eV seen both in absorption and luminescence.

We now turn to the mechanism of the SGL. The model proposed by Dingle [18] assumes that in the first step, after creation of electron-hole pairs, a photoelectron is captured by $Cu^{2+}$ on a shallow donor level. In the second step, the negatively charged $Cu^+$ ion binds a hole on a shallow acceptor level, with binding energy of about 0.4 eV. Finally, a radiative transition from $[Cu^{1+}(d^{10})$, hole] to $Cu^{2+}(d^9)$ configurations takes place. This model, together with the (0/-) level of Cu being 0.2 eV below the CBM [21, 29], explains the observed 2.86 eV energy of the SGL transition. Dahan *et al.* [20] performed detailed calculations based on this picture, and linked the 3-fold splitting of the SGL from the point (iii) with the 3-fold splitting of the shallow acceptor level, which reflects the splitting of the VBM in ZnO.

In the second model [33], the electron transition occurs between a shallow donor and Cu in the $Cu^{3+}(d^8)$ charge state, similarly to [22]. In this model, the observed small splitting of the ZPL (see point (iii)) is ascribed to the Jahn-Teller splitting of the $t_{2\downarrow}$ triplet gap state of Cu into three singlets. Justification of this picture comes from ab initio calculations, which predict such a splitting for $Cu^{3+}$, but not for $Cu^{2+}$. The model is problematic for three reasons. First, concentrations of $Cu^{3+}$ in ZnO are expected to be negligible given the strong tendency of this crystal to be *n*-type. Secondly, as it was mentioned, DAP lines are broad while SGL is sharp. Finally, according to Ref. [27] the first step of SGL is the electron capture by Cu, after which Cu would be in the 2+ rather than 3+ charge state.

The third proposed mechanism [38] consists in the transition of a photoelectron from the CBM to $Cu^{3+}$. The calculations reproduce well both the experimental SGL energy and the linewidth. The authors considered also the $(Cu^{1+}, hole) \rightarrow Cu^{2+}$ transition, obtaining a somewhat worst agreement regarding both the PL energy and the linewidth. The experiment, however, demonstrates that the initial state is $Cu^{1+}$, and the symmetry of the $Cu^{2+}$ is $C_{3v}$, and not $C_2$ obtained in [38], which questions the proposed mechanism.

Theoretical studies of isolated Cu in ZnO were performed using LDA/GGA [33, 39-41], LDA+*U* [36, 42, 43], and hybrid functionals (HY) [37, 38]. The underestimation of the host band gap $E_{gap}$ by LDA/GGA leads to wrong energies of Cu levels relative to the VBM and CBM, and therefore to wrong optical transition energies. Application of +*U* correction only to *d*(Zn) [42, 43] partially improved the situation. Finally, hybrid functionals [37, 38] render a correct $E_{gap}$, but interpretation of experiment was limited to the SGL. This situation justifies the present effort of making a link between theory and the experimental data.

Below, after presenting the computational method in Sec. II, we discuss Cu levels, formation energies and transition levels in Sec. III. Section IV is devoted to the interpretation of the PL optical data. Section V summarizes our results.

## 2. COMPUTATIONAL METHODS

Calculations based on the density-functional theory are performed using the generalized gradient approximation (GGA) [44], the Perdew-Burke-Ernzerhof exchange-correlation potential [45], and include the +*U* correction implemented in the QUANTUM-ESPRESSO code [46] along with the theoretical framework developed in Ref. [47]. Ultrasoft atomic pseudopotentials are employed, and the following valence orbitals were chosen: $3d^{10}$ and $4s^2$ for Zn, $2s^2$ and $2p^4$ for O, $3d^9$ and $4s^2$ for Cu. The plane wave basis with the kinetic energy cutoff of 40 Ry provides a convergent description of the analyzed properties. The Brillouin zone summations are performed using the Monkhorst-Pack scheme with a 2×2×2 *k*-point mesh [48]. Methfessel-Paxton smearing method with the smearing width of 0.136 eV is used to account for partial occupancies. Ionic positions are optimized until the forces acting on ions are smaller than 0.02 eV/Å. Cu atoms are placed in 72- and 128-atom supercells. Spin-orbit coupling is neglected.

From our previous calculations [49, 50] it follows that the +*U* terms should be applied to both *d*(Zn) and *p*(O) orbitals. The values $U_{Zn}$ =10 eV and $U_O$ = 7 eV reproduce both the experimental $E_{gap}$=3.4 eV and the energy of the *d*(Zn) band, centered about 8.1 eV below the VBM [51-53] in bulk ZnO [49]. The VBM is split by the wurtzite crystal field splitting into the doublet ($e_{VBM}$) and a singlet ($a_{VBM}$) lower in energy by 0.1 eV. The sensitivity of $E_{gap}$ to not only $U_{Zn}$ but also $U_O$ stems from the fact that the VBM as well as the minimum of the conduction band (CBM) contain a large contribution of O orbitals. With this choice of *U*s, theoretical lattice parameters *a* = 3.22 Å and *c* = 5.24 Å agree to within 1 % with the experimental data.

One technical aspect of calculations requires a comment. In the GGA+*U* approach, occupation matrices are used to specify the occupation of each of the *d*(Cu) orbitals. After choosing the initial occupation matrix two schemes of calculations are possible: in the first one the occupations are kept fixed, while in the alternative scheme they can vary during the calculations. It turns out that all possible initial occupations should be considered, and in many cases, especially for large $U_{Cu}$, the two procedures result in different final results. It is natural to choose the solution with the lowest energy, as it is discussed in Ref. [54]. In the present case, both schemes give the same final results for small $U_{Cu} \leq 2$ eV. However, for $U_{Cu} \geq 3$ eV, large qualitative differences appear between the two schemes. The fixed occupations scheme leads to the splitting of the Cu-induced "$t_{2\downarrow}$" triplet into a doublet and a singlet (as discussed in detail in Sec. III), which is absent within the varying occupation scheme. For $U_{Cu}$ =3 eV, the former approach provides the total energy $E_{tot}$ lower by 0.55 eV, and thus we consider this as the correct result. The doublet-singlet splitting increases with $U_{Cu}$, and it is close to that obtained with HY [35, 37], 2.7 eV, for large $U_{Cu} \approx 7$ eV. On the other hand, the usage of the varying occupation scheme leads to the vanishing spin polarization of $Cu^{2+}$ for $U_{Cu}$ >2 eV, as well as vanishing doublet-singlet splitting. The dependence of the total energy $E_{tot}$ on the occupation scheme is presented in Fig. 1.

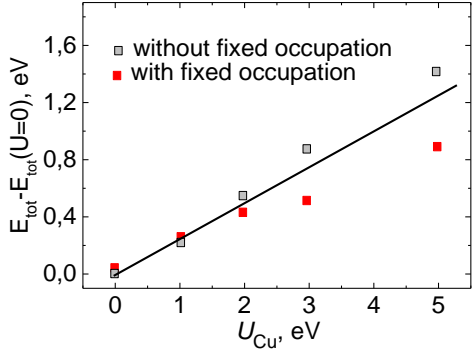

FIG. 1. Total energy as a function of $U_{Cu}$ with and without the fix occupation procedure, see text. Zero energy corresponds to $U_{Cu}=0$. The line is to guide the eye only.

To analyze probabilities of optical transitions we calculated squares of the momentum matrix elements. In the PW Quantum Espresso code, the wave functions are normalized by the conditions [46]:

$$<\psi_i | S | \psi_j > \equiv \begin{cases} 0, i \neq j \\ 1, i = j \end{cases}, \quad (1)$$

where S is the overlap matrix, which should be accounted for in the calculations. As a reference, in Table I we give the results for the three possible channels of interband CBM–VBM transitions, corresponding to the 3-fold degeneracy of the VBM. Since we are interested in relative intensities, in Tables I, III and IV we give the matrix elements of the gradient operator for the sake of simplicity. As expected, transition probabilities depend on the light polarization, which leads to selection rules: transition between the CBM and the VBM doublet is allowed for the $(x, y)$ polarization, while the transition between the CBM and the VBM singlet is allowed for the $z$ polarization. Note that the total probability of the CBM $− e_{VBM}$ transition for the $x$-polarization, 0.09+0.67=0.76, and for the $y$-polarization, 0.74+0.02=0.76, are equal.

Table I. Matrix elements squared of the gradient operator for the CBM→VBM interband transitions. $e_{VBM}(1)$ and $e_{VBM}(2)$ denote the two partners of the VBM doublet, while $a_{VBM}$ denotes the VBM singlet. The values are in atomic units (a.u.).

|       | $e_{VBM}(1)$ | $e_{VBM}(2)$ | $a_{VBM}$ |
|-------|--------------|--------------|-----------|
| $p_x$ | 0.09         | 0.67         | 0         |
| $p_y$ | 0.74         | 0.02         | 0         |
| $p_z$ | 0            | 0            | 0.84      |

## 3. ISOLATED Cu IN ZnO: ELECTRONIC STRUCTURE AND FORMATION ENERGIES

Local atomic configuration of Cu substituting for Zn is the same as that of the host Zn, *i.e.*, it has the $C_{3v}$ point symmetry. It is almost tetrahedral with a weak perturbation by the wurtzite geometry: the three planar Cu-O bonds are equivalent, and differ from the fourth bond parallel to the $z$-axis. In the wurtzite structure, the $d$(Cu) shell splits into a doublet $e$ and a higher in energy quasitriplet, denoted here by "$t_2$" (which is a triplet in the tetrahedral coordination). In the case of $Cu^{2+}$, those levels are located in the band gap. Unlike $Cu^{1+}$, which is a closed-shell system, both $Cu^{2+}$ and $Cu^{3+}$ are spin polarized. The energy gain due to spin polarization, defined as the difference in total energy of the spin-nonpolarized and spin-polarized calculations, is 1.1 and 1.6 eV, respectively, for $Cu^{2+}$ and $Cu^{3+}$, and $Cu^{3+}$ is in the $S=1$ high spin state.

The finite spin polarization is reflected in the exchange splitting $\Delta\varepsilon_{ex}$ of Cu levels into spin-up ("$t_{2\uparrow}$", $e_\uparrow$) and spin-down ("$t_{2\downarrow}$", $e_\downarrow$) states. The energies $\varepsilon$ of $e_\uparrow$, "$t_{2\uparrow}$", $e_\downarrow$, and "$t_{2\downarrow}$" relative to the VBM are 0.2 and 0.7, 0.71, and 1.4 eV, respectively (Fig. 2a). The magnetic moment of $Cu^{2+}$ of 1 $\mu_B$ is mainly located on the $d$(Cu) orbital (0.58 $\mu_B$), and partially on the $p$(O) orbitals of the four oxygen nearest neighbors (0.42 $\mu_B$), which follows from the projection of the total density of states onto the relevant atomic orbitals. Spin localization is also reflected in the spin density shown in Fig. 3 for both charge states.

The energies of Cu levels strongly depend on the charge state $q$. The levels of $Cu^{3+}$, $Cu^{2+}$ and $Cu^{1+}$, corresponding to charge states +1, 0, and -1, (*i.e.*, to $d^8$, $d^9$, and $d^{10}$ configurations) respectively, are shown in Fig. 2a. The changes in the level energies induced by the changes of the charge state are determined by the intracenter electron-electron coupling, which increases with the increasing number of electrons, and leads to an increase of level energies. Indeed, the change of $q$ from +1 to -1 leads to the increase of "$t_2$" by about 3 eV testifying the strong intrashell Coulomb repulsion. In the case of $q=+1$, the spin-up states are degenerate with the valence bands, while the spin down $e_\downarrow$ and "$t_{2\downarrow}$" are about 0.5 eV only above the VBM. Merging of the $Cu^{3+}$ spin up states into valence bands is accompanied with the pronounced hybridization of $Cu^{3+}$ states with valence bands, and with the delocalization of spin density, see Fig. 3. Finally, spin polarization of the negatively charged $Cu^+$ vanishes, and $e$ and "$t_2$" are about 0.8 and 0.3 eV below CBM, respectively.

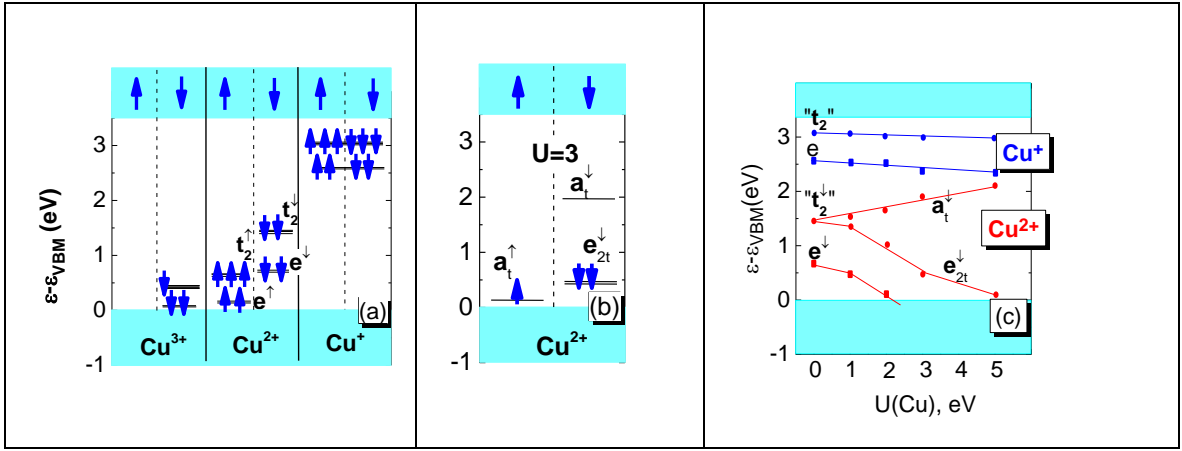

FIG. 2. (a) Energy levels of a Cu ion in ZnO for various charge states and $U_{Cu}=0$ eV. (b) Energy levels of $Cu^{2+}$ for $U_{Cu}=3$ eV. Note the large splitting of the "$t_{2\downarrow}$" triplet. Arrows and blue dots represent electron spins and empty states, respectively. Zero energy is at the VBM. (c) The levels of $Cu^{2+}$ and $Cu^{1+}$ in ZnO as function of $U_{Cu}$ assuming the fixed occupation scheme (see text).

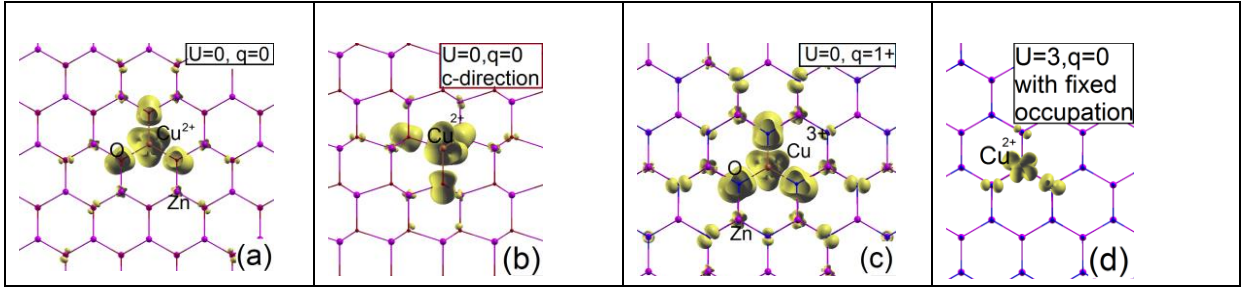

FIG. 3. Spin density of $Cu^{2+}$ (a, b, d), and of $Cu^{3+}$ (c) in ZnO. (a, c, d): the (0001) plane, and (b): the plane parallel to [0001]. Isosurfaces of spin densities correspond to 0.002 electron/bohr$^3$. (a-c): $U_{Cu}=0$, and (d) $U_{Cu}=3$ eV.

The impact of the $U_{Cu}$ term on the Cu levels is shown for $Cu^{1+}$ and $Cu^{2+}$ in Fig. 2c. Inclusion of $+U_{Cu}$ increases the energy difference between occupied and empty states. The corresponding energy correction is [47]:

$$\Delta\varepsilon_{m\sigma} = U(1/2 - n_{m\sigma}). \quad (2)$$

Here, $n_{m\sigma}$ is the occupation of a set of "localized" d(Cu) states with spin σ. If $|v\mathbf{k}\sigma\rangle$ is a ZnO band state with the wave vector **k** from the band v, then

$$n_{m\sigma} = \sum_{v\mathbf{k},\sigma} f(v\mathbf{k}\sigma) \langle v\mathbf{k}\sigma | m\sigma \rangle \langle m\sigma | v\mathbf{k}\sigma \rangle, \quad (3)$$

where f(v**k**σ) is the occupation number. In the case of $Cu^{1+}$, the dependence of both e and "$t_2$" on the $U_{Cu}$ value is unexpectedly weak, and leads to a decrease of Cu energies by about 0.2 eV only. We speculate that this non-sensitivity stems from the cancelation of two factors:

with the increasing $U_{Cu}$, the occupied d(Cu) levels are shifted downwards (according to Eq. (2)), but in parallel the localization of the d orbital increases, see Fig. 3, and so does the intrashell Coulomb repulsion, pushing the d(Cu) upward. Indeed, by comparing Figs. 3a and 3d one can see that the increase of $U_{Cu}$ increases the localization of Cu levels.

In contrast, in the case of $Cu^{2+}$, with the increasing $U_{Cu}$ corrections the spin-up states move down in energy, and eventually merge with the valence bands, and so does the $e_\downarrow$ doublet. In parallel, the splitting of the "$t_{2\downarrow}$" quasitriplet takes place into the occupied $e_{2t\downarrow}$ doublet and the empty $a_{t\downarrow}$ singlet. For $U_{Cu} = 3$ eV, the splitting is about 1.5 eV (Fig. 2b). We note that even a larger splitting, 2.7 eV, was obtained with HY in Ref. [37], which we reproduce with $U_{Cu} \approx 7$ eV.

Formation energy $E_{form}$ of a defect in a bulk material is given by [55, 56]:

$$E_{form} = E_{tot}(ZnO:Cu) - E_{tot}(ZnO) + \mu(Zn) - \mu(Cu) + q(E_F + E_{VBM}) \quad (4)$$

where the first two terms are the total energies of ZnO with and without the $Cu_{Zn}$, respectively. $E_F$ and $E_{VBM}$ are the Fermi energy and the VBM, and $q$ is the charge state of the impurity. $\mu(Zn)$ and $\mu(Cu)$ are the variable chemical potentials of Zn and Cu atoms in the solid, which in general are different from the chemical potentials $\mu(bulk)$ of the ground state of elements (Zn bulk, Cu bulk and $O_2$). Chemical potentials of the components in the standard phase are given by total energies per atom of the elemental solids: $\mu(Zn\ bulk) = E_{tot}(Zn\ bulk)$, $\mu(Cu\ bulk) = E_{tot}(Cu\ bulk)$, while $\mu(O\ bulk) = E_{tot}(O_2)/2$ that is the total energy per atom for $O_2$. In O-rich condition, $\mu(Zn) = E_{tot}(Zn\ bulk) + \Delta H_f(ZnO)$ and $\mu(Cu) = E_{tot}(Cu\ bulk) + \Delta H_f(CuO)$ are taken, where $\Delta H_f$ is the enthalpy of formation per formula unit, and it is negative for stable compounds. $\Delta H_f$ at T = 0 K is obtained by considering the reaction to form or decompose a crystalline ZnO and CuO from or into its components, and depends on cohesive energy $E_{coh}$ of Zn, Cu, and O. The obtained results for $E_{coh}$ of Zn, Cu, O, are 1.32 (1.35 [57]), 3.52 (3.49 [57]), and 2.85 (2.65 [57]), while $\Delta H_f(ZnO)$ = -3.7 (-3.6 [58]) and $\Delta H_f(CuO)$ = -1.62 (-1.59 [59]) eV, respectively (the experimental values are shown in the brackets). The calculated $E_{form}$ of Cu substituting Zn in ZnO in the O-rich conditions, and assuming $E_F$ at the VBM, are –0.33 eV for $Cu^{3+}$, 0.14 eV for $Cu^{2+}$, and 2.98 eV for $Cu^{1+}$, which is close to the results of Ref. [38], -0.36, 0.1, 2.9 eV, respectively.

The change in the impurity charge state is determined by the transition level $\varepsilon(q_1/q_2)$, defined as the Fermi energy relative to the VBM at which formation energies of the $q_1$ and $q_2$ charge states are equal. We find $\varepsilon(+/0) = 0.47$ eV and $\varepsilon(0/-) = 2.84$ eV, which is consistent with Cu energies shown in Fig. 2a. Comparable values of $\varepsilon(0/-)$ were obtained with different exchange-correlation functionals. For example, $\varepsilon(+/0) = 0.37$ (0.14) eV and $\varepsilon(0/-) = 2.32$ (3.46) eV is found in Ref. [36] without (with) additional hole-state corrections, $\varepsilon(0/-) = 2.48$ eV is obtained in Ref. [37], and $\varepsilon(+/0)=0.5$ eV and $\varepsilon(0/-) = 3.27$ eV is found in Ref. [38]. Atomic displacements around Cu affect $E_{form}$ and transition levels. The change of the charge state from $q=0$ to $-1$ is accompanied with the increase of Cu-O average bond lengths from 1.98 to 2.1 Å. When $U_{Cu} = 3$ eV, the three planar bonds $b_1$ are different from the remaining axial one $b_2$, and the change from $q=0$ to $-1$ is accompanied with the increase from $(b_1, b_2) = (1.92, 2.02)$ Å to $(1.97, 2.15)$ Å. The atomic displacements change the Cu levels by about 0.5 eV (or by 0.56 eV for $U_{Cu} = 3$ eV), and therefore is represents an important factor determining ionization energies.

## 4. Cu-RELATED OPTICAL TRANSITIONS

To interpret the experimental data, and in particular the internal transition of $Cu^{2+}$ at 0.72 eV and the SGL at 2.86 eV, we calculate both the transition energies given by the total energy differences between the excited and the ground state, and the corresponding matrix elements.

We first consider the internal transition. According to Fig. 2c, the absorption/recombination channel at 0.72 eV depends on the value of $U_{Cu}$. For $U_{Cu} < 2$ eV, in the initial excited state the "$t_{2\downarrow}$" triplet is fully occupied with 3 electrons and the $e_\downarrow$ doublet is occupied with one electron, while the final state is the $Cu^{2+}$ ground state. The calculated transition energies are given in Table II. Since the Cu charge state does not change, the corresponding atomic configurations are practically the same, and the relaxation energy is negligible, about 0.001 eV. When $U_{Cu}=0$ is assumed, the calculated transition energy, 0.72 eV, is in excellent agreement with experiment. Increasing $U_{Cu}$ to about 2 eV does not significantly affect this value. The situation deeply changes for $U_{Cu} > 2$ eV. In this regime, a large splitting of "$t_{2\downarrow}$" occurs, and $e_\downarrow$ eventually merges with the VBM, see Fig. 2. The intracenter transition may now occur between other levels, namely between the two "$t_{2\downarrow}$"-derived states, the $e_{2t\downarrow}$ doublet and the $a_{t\downarrow}$ singlet. However, the calculated energy of this transition does not exceed 0.62 eV (see Table II), which is smaller than the experimental value. To farther discriminate between these two possibilities we calculated the corresponding momentum matrix elements squared. There are 3×2=6 possible recombination channels for the "$t_{2\downarrow}$" → $e_\downarrow$ transition, 2×1= 2 for the $e_{2t\downarrow}$ → $a_{t\downarrow}$ transition, and each channel is polarization dependent. The results given in Table III clearly demonstrate that the probability of the $e_{2t\downarrow}$ → $a_{t\downarrow}$ case is about $10^{-5}$ smaller than that of the "$t_{2\downarrow}$" → $e_\downarrow$ transition. These results suggest the choice of $U_{Cu} < 2$ eV.

TABLE II. Calculated energies of the two internal transitions (see text) and the ($Cu^+$, hole) → $Cu^{2+}$ SGL transition for varying $U_{Cu}$. All values in eV. For $U_{Cu} = 2$ eV, two values corresponding to the two considered cases are given.

| $U_{Cu}$ | 0 | 1 | 2 | 3 | 5 |
|---|---|---|---|---|---|
| internal "$t_2$" → $e$ | 0.72 | 0.72 | 0.76 | | |
| internal $a_t$ → $e_{2t}$ | | | 0.35 | 0.56 | 0.62 |
| SGL | 2.86 | 2.85 | 2.73 | 2.7 | 2.61 |

Importantly, the above results show that energies of internal transition should not be estimated from the energy levels when the $+U$ corrections are applied. This is because the $d$(Cu) energies depend on occupation, see Eq. 2, which changes during a transition. This remark also applies to the results of HY calculations [37,38]. In this case, the calculated one-electron energies suggest that the $e_{2t\downarrow} \rightarrow a_{t\downarrow}$ internal transition occurs at about 2.5 eV, much larger than the experimental 0.72 eV, but a reliable conclusion requires the HY calculations of transition energy to be done.

TABLE III. Matrix elements squared of the gradient operator for the internal Cu transitions. The given values (in a.u.) are sums over the three polarizations and over the contributions of partners of the multiplet.

|  | $U_{Cu}$, eV | | | |
|---|---|---|---|---|
|  | 0 | 2 | 3 | |
|  | $t_2$ | $t_2$ | $a_t$ | $a_t$ |
| $e$ | 0.49 | 0.43 | | |
| $e_{2t}$ | | | ~$10^{-4}$ | ~$10^{-5}$ |

Regarding the transition responsible for the SGL, we follow the experiment indicating that it originates in the (Cu$^+$, hole) $\rightarrow$ Cu$^{2+}$ recombination of a $d$(Cu) electron with a free hole. Importantly, as it follows from Fig. 2c, there are two such possible transitions, because there are two $d$(Cu$^+$) levels below the CBM. Indeed, the recombining electron can initially reside either on the "$t_2$" or on the $e$ levels, but only the first case was analyzed in the literature so far. (Those two processes are denoted as "$t_2$" $\rightarrow$ VBM and $e \rightarrow$ VBM, respectively.) We note that both transitions could lead to photoluminescence, i.e., two rather than one line (or, more precisely, two sets of lines rather than one) should be observed. All possible transitions are schematically shown in Fig. 4. The calculated PL energies are given in Table II as a function of $U_{Cu}$. As in the case of the internal transition, $U_{Cu} < 2$ eV gives a very good agreement with experiment for the "$t_2$" $\rightarrow$ VBM process. The corresponding configuration diagram is displayed in Fig. 5. The calculated PL energy for the second channel, $e \rightarrow$ VBM, is lower by about 0.7 eV.

The reason for the apparent lack of the second PL line follows from the selection rules, which make one of the transitions less probable than the other. Because there are 5 levels of Cu$^+$ as initial locations of the recombining electron, "$t_2$" and $e$, and the VBM is split, there are 5×3=15 possible recombination channels to consider. Each of those is polarization dependent. The relevant levels are schematically shown in Fig. 4, and the calculated transition probabilities are given in Table IV. They show that the disparity between the two possible PL channels, "$t_2$" $\rightarrow$ VBM and $e \rightarrow$ VBM, stems from the fact that the latter process is less probable by over an order of magnitude. Moreover, the calculated matrix elements for the interband PL (Table I) and for the "$t_2$" $\rightarrow$ VBM are close, which is in accord with the observed efficiency of Cu as a radiative recombination center.

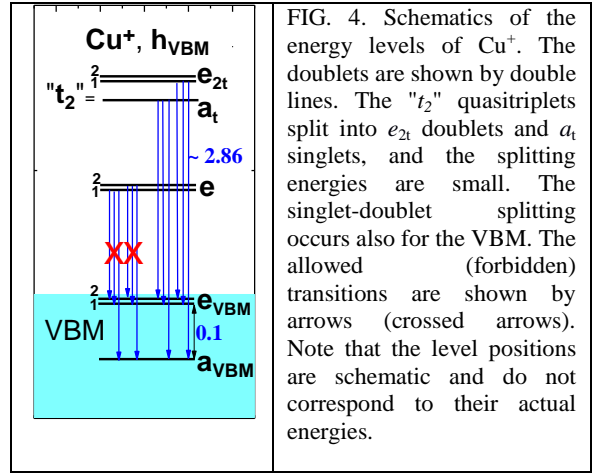

FIG. 4. Schematics of the energy levels of Cu$^+$. The doublets are shown by double lines. The "$t_2$" quasitriplets split into $e_{2t}$ doublets and $a_t$ singlets, and the splitting energies are small. The singlet-doublet splitting occurs also for the VBM. The allowed (forbidden) transitions are shown by arrows (crossed arrows). Note that the level positions are schematic and do not correspond to their actual energies.

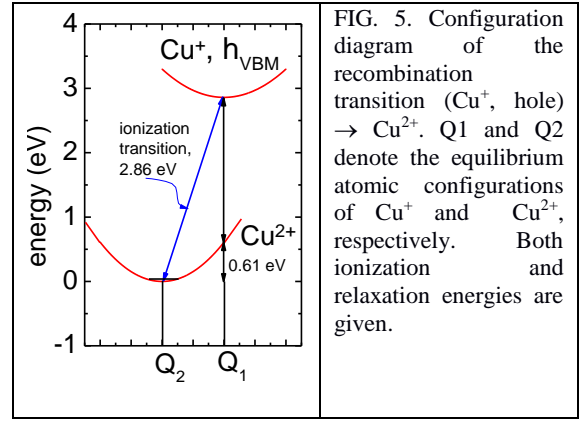

FIG. 5. Configuration diagram of the recombination transition (Cu$^+$, hole) $\rightarrow$ Cu$^{2+}$. Q1 and Q2 denote the equilibrium atomic configurations of Cu$^+$ and Cu$^{2+}$, respectively. Both ionization and relaxation energies are given.

Table IV. Matrix elements squared of the gradient operator for the $d$(Cu)-VBM ionization transitions. The given values (in a. u.) are sums over the three polarizations and over the contributions of two partners of doublets.

|  |  | Cu($e$) | Cu($t_2$) | |
|---|---|---|---|---|
|  |  |  | $a_t$ | $e_{2t}$ |
| VBM | $e_{VBM}$ | < 0.01 | 0.22 | 0.42 |
|  | $a_{VBM}$ | < 0.01 | 0.13 | 0.41 |

## 5. SUMMARY

The GGA+$U$ calculations were performed to provide interpretation of the experimental optical properties of Cu in ZnO. The calculations include the electronic structure of Cu, and energies of possible PL processes together with the corresponding transition probabilities. Formation energy and transition levels of Cu were also obtained.

The charge transfer transition leading to the structured green luminescence at 2.86 eV was analyzed. In the Dingle model [18], supported by a large number of experimental works reviewed in Introduction, the SGL follows from the recombination of a $d$ electron of $Cu^+$ with a free hole. Assuming ($Cu^+$, hole) as an initial configuration for the SGL, we obtain the ZPL luminescence energy of 2.84 eV, in excellent agreement with the measured value.

Next, theory shows that there are two $Cu^+$-induced levels in the band gap. Such a situation implies that a second recombination channel (from the lower $Cu^+$ gap state) can be expected, but only one PL line is observed. This feature is explained by the fact that the calculated transition probabilities of the second channel are much smaller than those of the SGL. Moreover, the calculated probabilities for the SGL are comparable to those for the interband PL, which is in accord with the observed efficiency of the SGL as the recombination channel.

We also analyzed the internal transition of $Cu^{2+}$ at 0.72 eV, which was observed in both PL and absorption, and interpreted as the internal $t_{2\downarrow}$-$e_\downarrow$ Cu transition. It was not previously considered by theory. Our calculated value agrees very well with the measurements. Also in this case the calculated transition probabilities allow distinguishing between the two possible origins (involving two different electronic configurations of $Cu^{2+}$) of the 0.72 eV line.

Technically, the +$U$ corrections were treated as fitting parameters, and applied to $d$(Zn), $p$(O) and $d$(Cu) orbitals. For both the SGL and the internal transitions, a good agreement with the experimental values is obtained when $U_{Cu} < 1.5$ eV are employed. For $U_{Cu} > 4$ eV, we obtain results comparable to those of the HY calculations. Namely, in the case of $Cu^{2+}$, the gap "$t_{2\downarrow}$" quasitriplet, occupied with 2 electrons, is split into the occupied $e_{2t\downarrow}$ doublet and the empty $a_{t\downarrow}$ singlet. The large splitting of 2.7 eV [37] is acompanied with the lowering of the point symmetry from $C_{3v}$ to $C_2$ [38], in which the three planar bonds differ by 7 %. The effect was dubbed the pseudo-Jahn-Teller effect [60], because the splitting of the gap multiplet is related with the electron-electron coupling, see Eq. 2, and it can take place even without the lowering of the defect point symmetry by atomic displacements, *i.e.*, without the Jahn-Teller effect. However, the results are in contradiction with experiment regarding the local symmetry of the Cu ions.

**Acknowledgements**. We thank L. Klopotowski for helpful discussions. The work was supported by the National Science Centre, (Poland) Grant No. 2015/17/D/ST3/00971. Calculations were done at Interdisciplinary Center for Mathematical and Computational Modeling, University of Warsaw (Grant No. GA65-27).